# Experimental comparison of various techniques for spot size measurement of high-energy x-ray source

Yi Wang, Qin Li, Nan Chen, Jinming Cheng, Chenggang Li, Hong Li, Quanhong Long, Jinshui Shi, Jianjun Deng

Key Laboratory of Pulsed Power, Institute of Fluid Physics, China Academy of Engineering Physics, Mianyang 621900, Sichuan Province, China

**Abstract:** In the experiment of flash-radiography, the quality of acquired image strongly depends on the focal size of the x-ray source spot. A variety of techniques based on imaging of pinhole, slit and rollbar are adopted to measure the focal spot size of the Dragon-I linear induction accelerator. The image of pinhole provides a two-dimensional distribution of the x-ray spot, while those of slit and rollbar give a line-spread distribution and an edge-spread distribution, respectively. The spot size characterized by full-width at half-maximum and that characterized by the LANL definition are calculated for comparison.

## 1. Introduction

In the experiment of high-energy flash radiography, the x-ray source comes from the bremsstrahlung radiation generated by accelerating, transporting and focusing an electron beam pulse onto a heavy-metal target. [1-5] The quality of acquired image is closely related to the size of the source spot, which is often quoted as an evaluation of the resolving power of a particular flash-radiography machine. The focal spot size of the x-ray source strongly depends on the electron beam size and the scattering of electrons and photons within the target. A good knowledge of the size and the shape of the source spot is of great importance not only to the inversion of the material density but also to the design optimization of the target.

It is not an easy task to precisely measure the focal size of a radiographic source which has a very high energy (~MeV) due to a strong transmission of material. A number of techniques have been proposed for high-energy x-ray spot size measurement, which utilizes a pinhole [6], a slit [7] or a rollbar [8] for imaging. These methods provide different information of the source spot. A full two-dimensional spatial distribution of the x-ray spot can be obtained by the pinhole method. The images of the slit and the rollbar actually denote a line-spread function (LSF) and an edge-spread function (ESF), respectively. Various definitions of spot size have also been introduced, which characterizes the spot size from different aspects. [9-11] For example, the full-width at half-maximum (FWHM) of the spot simply considers a specific boundary of the spatial distribution while the LANL definition [9] involves the spatial frequency of the modulation transfer function (MTF). In this paper, we apply different techniques to measurement the x-ray spot size of the Dragon-I linear induction accelerator (LIA) [5]. The results of spot size based on both the FWHM and the LANL definition are given for comparison.

## 2. Principle

After emitting from the x-ray source, photons pass through the object placed in the light field and finally reaches the receiving system for image recording. According to the transfer property of

the optical function, the linear process follows the relation of
$$i(x, y) = s(x, y) * o(x, y) * r(x, y), \quad (1)$$
where $i(x, y)$ is the final recorded image of the system; $s(x, y)$ is a two-dimensional spatial distribution of the x-ray source, i.e. the point-spread function (PSF); $o(x, y)$ represents the transfer property of the object, which corresponds to point source imaging; $r(x, y)$ is the blur of the image-receiving system; the sign, $*$, denotes the convolution operation.

By making Fourier transform of Eq. (1), the relation of modulation transfer function (MTF) of the system can be expressed as
$$I(f) = S(f) \cdot O(f) \cdot R(f), \quad (2)$$
where $I(f)$, $S(f)$, $O(f)$ and $R(f)$ represent the MTF of each term in Eq. (1). If the transfer property of the object can be expressed by the delta function and the blur of image-receiving system be ignored, the relation will be simplified as
$$I(f) \approx S(f). \quad (3)$$
In this condition, the transfer property of the obtained image is exact a reflection of the x-ray source.

The setup of spot size measurement utilizing the pinhole or the slit imaging technique is illustrated in Fig. 1. A pinhole or slit object is placed between the conversion target and the image receiving plane. The diameter of the pinhole (or the width of the slit) is denoted as $d$, with an axial thickness of $L$. The distances from the rear side of the object to the source plane and the image screen are denoted as $a$ and $b$, respectively. Then the geometrical magnification is defined as $M = b/a$. Following the principle of geometrical similarity, the relation between the source size and the image size can be written as
$$D_0 = D/M - (1 + 1/M)d, \quad (4)$$
where $D_0$ and $D$ represent the sizes of source and image, respectively. It should be noticed that the sizes mentioned in Eq. (4) actually denote the boundary of edge. But in practice, the light source generally has an expanding spatial distribution, the boundary of which is probably unclear. In the circumstances, Eq. (4) is inappropriate to describe the relation of the source size and the image size characterized by FWHM. From Eq. (3), the focal spot size can be simply calculated as
$$D_0 = D/M. \quad (5)$$

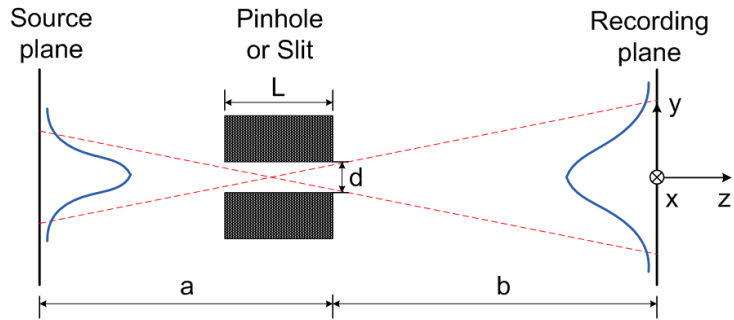

Fig. 1 Sketch of spot size measurement by pinhole or slit imaging method.

The setup of roll bar spot size measurement is drawn in Fig. 2. Instead of using opaque shield with a knife-edge, a thick heavy-metal bar with a roll curving edge (the curvature radius is denoted as $R$) is employed to ease alignment. Due to a partial block of the radiation from the source, the obtained penumbral image directly yields the ESF, which is actually the convolution of the LSF with a step function centered at the edge. Therefore, the LSF can be calculated through differentiating the ESF along the direction perpendicular to the edge. Then the source FWHM is obtained by dividing the magnification to the image FWHM.

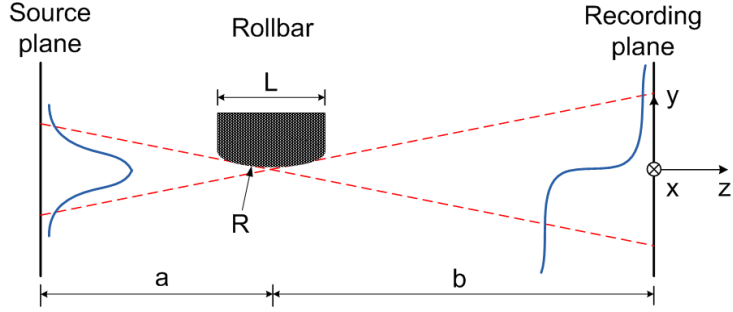

Fig. 2 Sketch of spot size measurement by rollbar method.

The spot size of FWHM takes into consideration the boundary of a particular level rather than the spatial distribution. In LANL definition, the spot size ($D_{LANL}$) is characterized from the spatial frequency. The MTF obtained by making Fourier transform of LSF analyzes each imaging component as a low pass filter for spatial information. Practically, $D_{LANL}$ is defined as the diameter of an equivalent uniform disk that has the same spatial frequency at half of the MTF peak value [9]:

$$D_{LANL} = \frac{0.705}{f_{50\% \, MTF} \cdot M}, \qquad (6)$$

where $f_{50\% \, MTF}$ is the spatial frequency of the image LSF, with the unit of inverse dimension.

## 3. Experiments

Experiments are performed on the Dragon-I LIA, which is able to generate a high-energy, intense-current electron beam pulse. The x-ray source is produced from bremsstrahlung radiation by transporting and focusing the electron beam onto a target of tantalum layer with the thickness of 1.2 mm. For the purpose of maintaining a stable x-ray source, the energy of electron beam is kept in the range of $E = 18.8 \sim 19.1 \, \text{MeV}$, and the current in the range of $I = 2.46 \sim 2.48 \, \text{kA}$. Besides, the currents loaded on the solenoids in the downstream section for transporting and focusing the electron beam remain invariant during the measurements. The x-ray imaging system consists of a screen made of CsI scintillator, a flat mirror tilted at $45°$ angle with respect to the beam direction, and a CCD camera.

Firstly, a tungsten cube object with a pinhole through is placed between the x-ray source and the scintillating screen. The diameter of the pinhole is $d = 0.47 \, \text{mm}$, and its axial thickness is $L = 65 \, \text{mm}$. The distances of the setup are $a = 1190 \, \text{mm}$, $b = 5075 \, \text{mm}$, which indicate a magnification factor of $M = 4.26$. The obtained spot image in the experiment can be projected in both the horizontal and the vertical directions to provide two LSFs, each of which follows a direction perpendicular to the projection.

Then the slit method is applied for the spot size measurement. The material of the slit object is tungsten. The width and the thickness of the slit are $d = 0.30 \, \text{mm}$ and $L = 45 \, \text{mm}$, respectively. The parameters of the setup are as follows $a = 1181 \, \text{mm}$, $b = 5084 \, \text{mm}$ and $M = 4.30$. The slit is laid horizontally (in x-direction), which gives a LSF along the y-direction.

After that, a tungsten rollbar is used for x-ray spot measurement. The parameters of the rollbar and the experimental alignment are given as follows: $L = 120$ mm, $R = 1000$ mm, $a = 1164$ mm, $b = 5101$ mm and $M = 4.38$. The rollbar is also laid horizontally, so the distribution along the vertical direction of the shadow provides the ESF of the acquired image.

## 4. Results and discussions

For each technique, three measurements are taken to measure the FWHM of the x-ray source, the results of which are listed in Table 1. The pinhole imaging method can provide not only the PSF FWHM of the source but also the LSF FWHMs corresponding to projections of two orthogonal directions. The slit and the rollbar methods are only able to acquire the FWHM of the LSF along the y-direction due to the experimental alignment of the slit or the rollbar object.

A typical spot image obtained by the pinhole imaging technique (No. 13031) is shown in Fig. 3, where the black curve denotes the boundary of 50% peak value and the white curve denotes that of 10% peak value. The FWHM of the spot image is given by the diameter of a circular disk that has the same area as the inside of the 50%-peak-value contour. Using Eq. (5), the FWHM of the source PSF is figured out to be 1.32 mm. Two LSF curves along the x-direction and the y-direction are obtained corresponding to the vertical and the horizontal projections, each of which has an FWHM of 1.75 mm and 1.71 mm, respectively. Fig. 4 shows a typical image of slit spot size measurement (No. 13041). The vertical intensity distribution stands for the LSF along that direction, which finally gives rise to an LSF(y) FWHM of 1.59 mm for the source spot. In Fig. 5, the penumbral image (No. 13044) acquired by the rollbar method denotes the ESF, the first derivative of which is the LSF along the y-direction. The FWHM of LSF(y) on the focal plane is 1.63 mm. Moreover, each LSF curve of the acquired image is compared with theoretical functions with a same FWHM, including Gaussian (GS), Bennett (BNT) and Quasi-Bennett (QBNT) distributions [11].

Table 1 Experimental results of spot size characterized by FWHM.

| Method | No. | M | E / MeV | I / kA | FWHM of PSF / mm | FWHM of LSF(x) / mm | FWHM of LSF(y) / mm |
|---|---|---|---|---|---|---|---|
| Pinhole | 13031 | 4.26 | 18.9 | 2.46 | 1.32 | 1.75 | 1.71 |
|  | 13034 | 4.26 | 19.1 | 2.47 | 1.29 | 1.68 | 1.71 |
|  | 13035 | 4.26 | 19.0 | 2.48 | 1.34 | 1.67 | 1.75 |
| Slit | 13041 | 4.30 | 18.9 | 2.46 | \ | \ | 1.59 |
|  | 13042 | 4.30 | 18.9 | 2.46 | \ | \ | 1.67 |
|  | 13043 | 4.30 | 18.9 | 2.46 | \ | \ | 1.60 |
| Rollbar | 13044 | 4.38 | 19.1 | 2.47 | \ | \ | 1.63 |
|  | 13045 | 4.38 | 19.0 | 2.47 | \ | \ | 1.67 |
|  | 13047 | 4.38 | 18.8 | 2.47 | \ | \ | 1.57 |

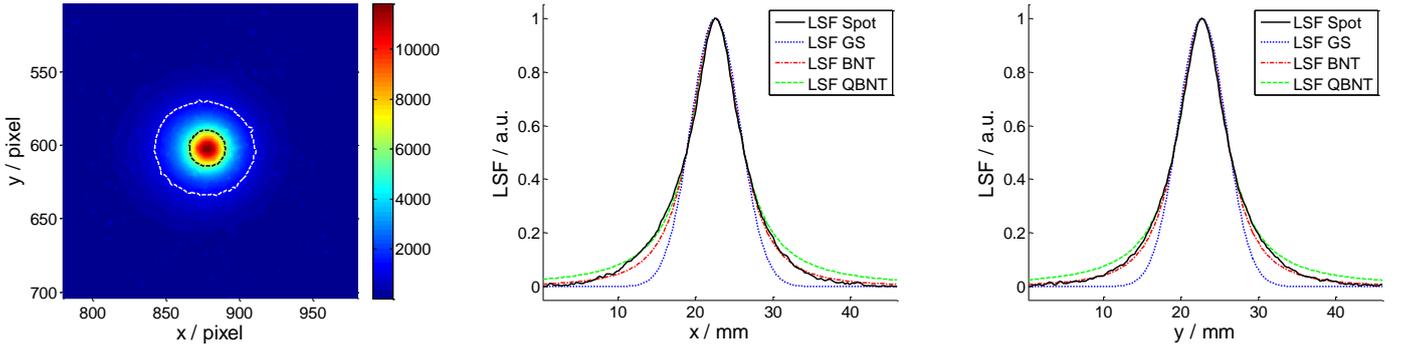

Fig. 3 Typical spot image and LSF curves by the pinhole imaging method (No. 13031).

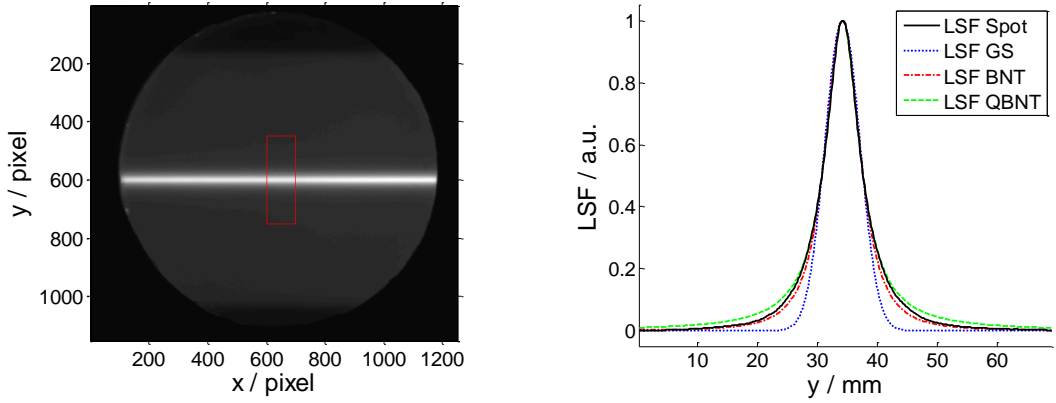

Fig. 4 Typical slit image and LSF curve by the slit method (No. 13041).

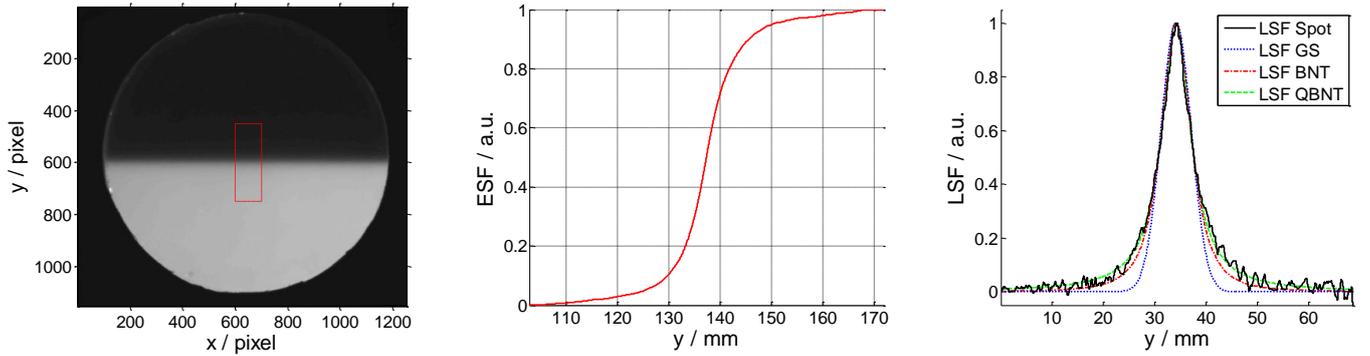

Fig. 5 Typical rollbar image and the ESF and LSF curves by the rollbar method (No. 13044).

For each experiment, the spot size in LANL definition is calculated. The $\mathrm{MTF}(f_x, f_y)$ is a two-dimensional surface in the spatial frequency space. According to the projection-slice theorem, the Fourier transform of the projection of the PSF in a direction (here take x-direction projection for example) yields to a slice ($f_x = 0$) of the MTF along the orthogonal direction, i.e.

$$\mathrm{MTF}(0, f_y) = \left| \int_{-\infty}^{+\infty} dy \int_{-\infty}^{+\infty} \mathrm{PSF}(x, y) \exp\left(-i 2\pi f_y y\right) dx \right|$$

$$= \left| \int_{-\infty}^{+\infty} \text{LSF}(y) \exp\left(-i2\pi f_y y\right) \text{d}y \right|. \tag{7}$$

After finding the spatial frequency at half of the MTF maximum, the LANL spot size is calculated by Eq. (6). In order to make a deduction of the screen blur, a 10-mm-thick tungsten plate with a straight edge is placed in contact with the scintillator, by which the ESF of the screen blur is obtained. The following treatment of the blur data is similar to that of the rollbar technique. The results of the LANL spot size for different techniques are listed in Table 2. Typical MTF results of the images are drawn in Fig. 6, each of which contains the MTFs of the image LSF, the screen blur and the LSF with blur deducted.

It is seen that the results of $D_{\text{LANL}}$ are all greater than the FWHM of the source PSF. In fact, $D_{\text{LANL}}$ defined from spatial frequency and MTF has an intrinsic relation with the spatial distribution of the source spot, which is expected to become greater when the wing of the spatial distribution, with a same FWHM, expands broader. For a GS distribution, $D_{\text{LANL}}$ is 1.6 times the FWHM of PSF in theory. This theoretical ratio becomes 2.7 for a BNT distribution and 4.1 for a QBNT distribution. [11] For each method, the experimental ratio of $D_{\text{LANL}}$ to FWHM of PSF can be worked out. The average ratios are calculated to be 2.7 for the pinhole method, 2.6 for the slit method and 2.8 for the rollbar method, which denote a spatial distribution of the source spot close to BNT. It also can be verified by the experimental LSF curves, which show to be more likely a BNT distribution than other theoretical functions.

Table 2 Experimental results of spot size characterized by LANL definition.

| Method | No. | $f_{x,50\%\text{MTF}}$ / mm$^{-1}$ | | $D_{x,\text{LANL}}$ / mm | $f_{y,50\%\text{MTF}}$ / mm$^{-1}$ | | $D_{y,\text{LANL}}$ / mm |
|---|---|---|---|---|---|---|---|
| | | with blur | without blur | | with blur | without blur | |
| Pinhole | 13031 | 0.0391 | 0.0450 | 3.67 | 0.0402 | 0.0471 | 3.51 |
| | 13034 | 0.0389 | 0.0451 | 3.67 | 0.0398 | 0.0467 | 3.54 |
| | 13035 | 0.0405 | 0.0475 | 3.48 | 0.0402 | 0.0473 | 3.50 |
| Slit | 13041 | \ | \ | \ | 0.0408 | 0.0487 | 3.36 |
| | 13042 | \ | \ | \ | 0.0392 | 0.0462 | 3.54 |
| | 13043 | \ | \ | \ | 0.0397 | 0.0467 | 3.51 |
| Rollbar | 13044 | \ | \ | \ | 0.0358 | 0.0421 | 3.82 |
| | 13045 | \ | \ | \ | 0.0372 | 0.0440 | 3.66 |
| | 13047 | \ | \ | \ | 0.0378 | 0.0452 | 3.56 |

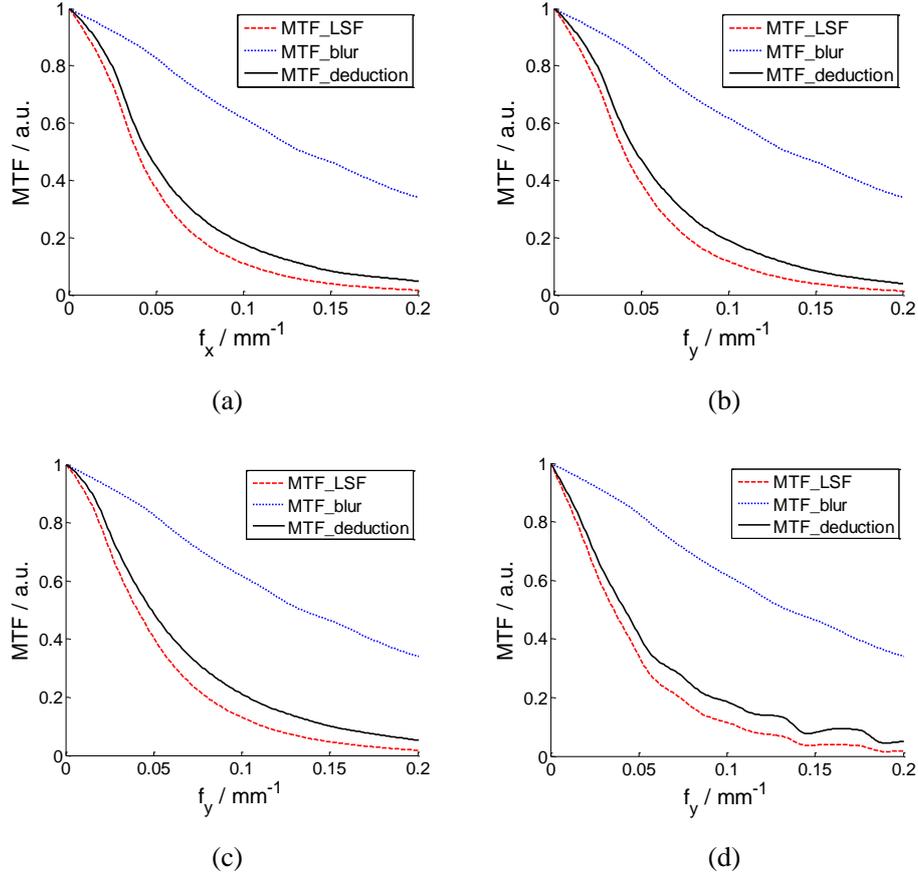

Fig. 6 MTF curves obtained in spot size measurements by various techniques. (a) and (b) Pinhole imaging method, No. 13031; (c) slit method, No. 13041; (d) rollbar method, No. 13044.

## 5. Conclusion

Different methods involving imaging of a pinhole, a slit and a rollbar are adopted to measure the focal size of the x-ray source spot of the Dragon-I LIA. The pinhole imaging technique provides a two-dimensional spatial distribution of the x-ray spot. The diameter of a disk that has the same area as the contour at half-peak-maximum gives the FWHM of the PSF, which is measured to be 1.32 mm on average. The acquired image of slit denotes an LSF along the direction orthogonal to the slit edge. The image of rollbar denotes an ESF along the direction orthogonal to the rollbar edge, the first derivative of which is the LSF. The LANL spot size is calculated for each method, the average value of the LANL spot size is obtained to be 3.56 mm by the pinhole method, 3.47 mm by the slit method and 3.68 mm by the rollbar method. The ratio values of $D_{\text{LANL}}$ to PSF FWHM with different techniques indicate that the x-ray source spot in the experiment is close to a BNT distribution.